\documentclass[aps,prl,showpacs,amsmath,amssymb,twocolumn]{revtex4-1}
\pdfoutput=1

\usepackage{color}
\usepackage{amsmath}
\usepackage{graphicx}
\usepackage{multirow}
\usepackage{float}

\begin{document}

\newcommand{\bn}{{\bf n}}
\newcommand{\bp}{{\bf p}}   
\newcommand{\br}{{\bf r}}
\newcommand{\bk}{{\bf k}}
\newcommand{\bv}{{\bf v}}
\newcommand{\brho}{{\bm{\rho}}}
\newcommand{\bj}{{\bf j}}
\newcommand{\wk}{\omega_{\bf k}}
\newcommand{\nk}{n_{\bf k}}
\newcommand{\eps}{\varepsilon}
\newcommand{\la}{\langle}
\newcommand{\ra}{\rangle}
\newcommand{\be}{\begin{eqnarray}}
\newcommand{\ee}{\end{eqnarray}}
\newcommand{\intl}{\int\limits_{-\infty}^{\infty}}
\newcommand{\dE}{\delta{\cal E}^{ext}}
\newcommand{\SE}{S_{\cal E}^{ext}}
\newcommand{\dsp}{\displaystyle}
\newcommand{\phit}{\varphi_{\tau}}
\newcommand{\p}{\varphi}
\newcommand{\cL}{{\cal L}}
\newcommand{\dphi}{\delta\varphi}
\newcommand{\dbj}{\delta{\bf j}}
\newcommand{\dI}{\delta I}
\newcommand{\dph}{\delta\varphi}

\title{How an incorrect transition from finite to infinite 2D conductor may result in a false
  negative relaxation}
  
\author{K.~E.~Nagaev}
\affiliation{Kotelnikov Institute of Radioengineering and Electronics, Mokhovaya 11-7, Moscow, 125009 Russia}

\date{\today}

\begin{abstract}
We consider the relaxation of a uniform current in a planar 2D conductor with account taken of 
electromagnetic retardation effects. If the 2D conductivity is larger than the speed of light, the
straightforward solution for an infinite plane gives a negative relaxation rate. However if one starts 
from a conducting cylinder of finite radius and then increases it to infinity, the relaxation rate just tends to zero while remaining positive. We suggest that recent unusual plasmon-dispersion curves obtained by 
V.~A.~Volkov and A.~A.~Zabolotnykh  [arXiv:1605.00430] result from the incorrect finite-to-infinite transition.
\end{abstract}

\maketitle

The charge dynamics in two-dimensional conductors is understood rather well in the absence of dissipation. The
dispersion law of plasmons in an infinite 2D plane with account taken of retardation effects was established by Stern \cite{Stern67} in
the absence of magnetic field. The dispersion curve consisted of a single branch that lied below the light line
$\omega=ck$. Several years later, Chiu and Quinn \cite{Chiu74} considered the same problem in a strong magnetic field perpendicular to the plane of 2D conductor. They also obtained a single-valued $\omega(k)$ dependence, 
which yet  exhibited discontinuities at multiples of the cyclotron frequency. The problem becomes much more complicated if the dissipation in the conductor is taken into account. The authors of \cite{Falko89}
took into account the dissipation in the infinite conducting plane within the Drude model and obtained that the
plasmon dispersion essentially depends on the relation between the 2D dc conductance of the plane $\sigma_0$ and
the speed of light $c$. According to \cite{Falko89}, the dispersion curve has a branching point at finite $k$ if  $2\pi\sigma_0<c$ and is single-valued if $2\pi\sigma_0>c$. They stated that in the latter case, the group velocity 
of plasmons $\partial\omega/\partial k$ could exceed the speed of light. 

In a very recent paper \cite{Volkov16}, 
the 2D plasmon dispersion was addressed in a presence of both dissipation and magnetic field normal to the plane. The authors
obtained a complicated pattern of dispersion curves that depends on the ratio $2\pi\sigma_0/c$ and the product
of the cyclotron frequency and the Drude relaxation time. For some values of these parameters, the imaginary part
of $\omega$ became positive at a finite $k$, while the real part of $\omega$ remained nonzero. This suggested an
emergence of instability  and   spontaneous oscillations with wave vectors above this value of $k$
in the absence of external sources of energy. The authors interpreted the change of sign of relaxation rate as a termination of the dispersion curve at this point, but we suggest a different explanation of this controversial result. 
In our opinion, the negative relaxation obtained in \cite{Volkov16} may result from the incorrect limiting 
transition from
finite to infinite conductor. Below we illustrate this point by a very simple example  of uniform current relaxation
in a 2D conductor.

Consider a 2D conductor with electron scattering that carried a uniform current at the initial moment. If no 
external fields are applied to it, the current will decay with time.
Like in Refs. \onlinecite{Falko89} and \onlinecite{Volkov16}, we assume that the 2D density of current is described
by the Drude formula  
\be
 \left( \frac{\partial}{\partial t} + \frac{1}{\tau}\right) {\bf j}_2 
 = \frac{\sigma_0}{\tau}\,{\bf E}_{\parallel},
 \label{Drude}
\ee
where $\tau$ is the momentum relaxation time of electrons, $\sigma_0 = e^2 n_2 \tau/m$ is the 2D conductance,
and ${\bf E}_{\parallel}$ is the component of electric field parallel to the 2D conductor. The total 
electric field is expressed in terms of the scalar and vector potentials as
\be
 {\bf E} = -\frac{1}{c}\,\frac{\partial{\bf A}}{\partial t} - {\bm\nabla}\p,
 \label{E-gen}
\ee
and the vector and scalar potentials obey the equations
\begin{align}
 \nabla^2{\bf A} - \frac{1}{c^2}\,\frac{\partial^2{\bf A}}{\partial t^2}
 = - \frac{4\pi}{c}\,{\bf j},
 \label{A-gen}\\
 \nabla^2\p - \frac{1}{c^2}\,\frac{\partial^2\p}{\partial t^2}
 = - 4\pi\rho.
 \label{p-gen}
\end{align}
Because we consider the uniform relaxation of current,  $\bm\nabla{\bf j}=0$, and therefore $\rho=0$, so
the last term in Eq. \eqref{E-gen} is zero. We expect all the relevant quantities to be proportional to 
$\exp(-\nu t)$, where $\nu$ is the inverse relaxation time to be determined. Therefore the system of equations 
for the current density and vector potential reduces to
\begin{align}
 (-\nu + 1/\tau)\,{\bf j}_2 &= \frac{\nu\sigma_0}{c\tau}\,{\bf A}_{\parallel},
 \label{j-rel}\\
 \nabla^2{\bf A} - \frac{\nu^2}{c^2}\,{\bf A} &= - \frac{4\pi}{c}\,{\bf j}.
 \label{A-rel}
\end{align}

First we assume that the 2D conductor is an infinite plane with the  normal to the it in the $z$ direction. 
Hence one may substitute ${\bf j} = {\bf j}_2\,\delta(z)$ into Eq. \eqref{A-rel}, and the problem becomes 
purely one-dimensional. The solution of Eq. \eqref{A-rel} is of the form
\be
 {\bf A}(z)  = \frac{2\pi}{\nu}\,{\bf j}_2\,e^{-\nu|z|/c}.
  \label{A-plane-3}
\ee
\begin{figure}[H]
 \includegraphics[width=8.5cm]{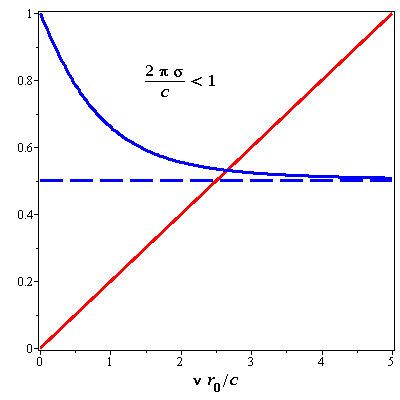}
 \caption{Graphic solution of Eq. \eqref{nu-cyl-eq} for $\tau c/r_0 =1/5$ and $2\pi\sigma_0/c=1/2$.}
 \label{fig1}
\end{figure}\noindent
A substitution of ${\bf A}(0)$ into Eq. \eqref{j-rel} readily gives the relaxation rate
\be
 \nu = \frac{1}{\tau}\left( 1 - \frac{2\pi\sigma_0}{c} \right).
 \label{nu-plane}
\ee

\noindent
This solution was presented in Ref. \onlinecite{Falko89} for $2\pi\sigma_0 <c$, where it is positive.
However it was not mentioned there that it also exists in the case of opposite inequality, which implies
an unphysical amplification of the initial current.

To clarify the origin of the negative relaxation rate, we note that a similar system of a small size represents
just an $RL$ circuit, in  which this rate is positive  regardless of the conductance. Therefore it makes
sense to start with a 2D conductor of finite size and trace the behaviour of the relaxation rate as its size
increases. Consider a conductor in the shape of an infinitely long hollow cylinder of radius $r_0$ with a 
uniform current flowing normally to the axis of the cylinder. Because of the symmetry of the problem, all quantities
depend only on the radial distance from this axis, and the vector potential has only the
azimuthal component. In the cylindrical coordinate system, Eq. \eqref{A-rel} takes up the form
\be
 \frac{d}{dr} \left[\frac{1}{r}\,\frac{d}{dr}\,(rA_{\p})\right]
 -\frac{\nu^2}{c^2}\,A_{\p}
 =-\frac{4\pi}{c}\,j_2\,\delta(r-r_0).
 \label{A-cyl-eq}
\ee
The solution to this equation that is finite at $r=0$ and 
tends to zero at $r\to\infty$ may be written as
\be
 A_{\p}(r) = a
 \begin{cases}
  K_1(\nu r_0/c)\,I_1(\nu r/c), & r<r_0
  \\
  I_1(\nu r_0/c)\,K_1(\nu r/c), & r>r_0,
 \end{cases}
 \label{A-cyl-sol}
\ee
\begin{figure}[H]
 \includegraphics[width=8.5cm]{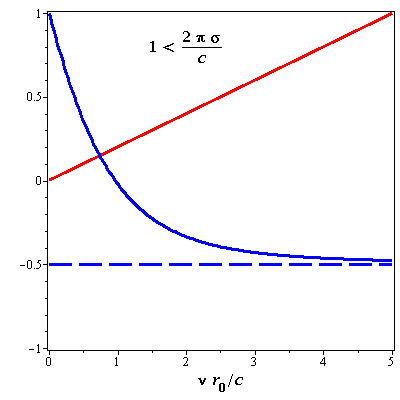}
 \caption{Graphic solution of Eq. \eqref{nu-cyl-eq} for $\tau c/r_0 =1/5$ and $2\pi\sigma_0/c=3/2$.}
 \label{fig2}
\end{figure} \noindent
where $I_1$ and $K_1$ are the modified Bessel functions of the first and second kind and the 
coefficient $a$ equals
\begin{multline}
 a = \frac{4\pi j_2}{\nu}\, \bigl[I_0(\nu r_0/c)\, K_1(\nu r_0/c) 
 \\
 + I_1(\nu r_0/c)\, K_0(\nu r_0/c)\bigr]^{-1}.
 \label{a}
\end{multline}
\noindent
A substitution of Eq. \eqref{A-cyl-sol} into \eqref{j-rel} gives the equation for $\nu$ in the form
\be
 \nu\tau = 1 - \frac{4\pi\sigma_0}{c}\,{\cal F}(\nu r_0/c),
 \label{nu-cyl-eq}
\ee
where
\be
 {\cal F}(x) \equiv \frac{I_1(x)\, K_1(x)}{I_0(x)\, K_1(x) + I_1(x)\, K_0(x)}.
 \label{F}
\ee
This function may be approximated as 
\be
 {\cal F}(x) \approx
 \begin{cases}
  x/2, & x \ll 1
  \\
  1/2, & x \gg 1,
 \end{cases}
 \label{Bessel-approx}
\ee
so the limiting forms of the solution to Eq. \eqref{nu-cyl-eq} are
\be
 \nu \approx
 \begin{cases}
  \dfrac{1}{\tau} \left(1 + \dfrac{2\pi\sigma_0}{c}\,\dfrac{r_0}{c\tau} \right)^{-1},
  & \dfrac{c\tau}{r_0} + \dfrac{2\pi\sigma_0}{c} \gg 1
  \\{} & {}\\
  \dfrac{1}{\tau}\left( 1 - \dfrac{2\pi\sigma_0}{c} \right),
  & \dfrac{c\tau}{r_0} + \dfrac{2\pi\sigma_0}{c} \ll 1.
 \end{cases}
 \label{nu-cyl-sol}
\ee
Therefore in the cylindrical geometry, the relaxation rate never becomes negative, no matter how large $r_0$ is.
If $2\pi\sigma_0/c>1$, the relaxation rate tends to zero with increasing $r_0$. This suggests that in 
a presence of dissipation, the transition from finite-size to infinite  2D systems is nontrivial and should be 
made with caution. 

The solution of Eq. \eqref{nu-cyl-eq} may be illustrated by Figures \ref{fig1} and \ref{fig2}, where the red lines represent
linear functions $(\tau c/r_0)\,x$ and the solid blue curves represent $1 - (4\pi\sigma_0/c)\,{\cal F}(x)$. The dashed blue lines show the asymptotic values of $1 - (4\pi\sigma_0/c)\,{\cal F}(\infty)$. The intersections of the red
and solid blue lines represent the solutions of Eq. \eqref{nu-cyl-eq} for the hollow cylinder, and their 
intersections with dashed blue lines present the values of $\nu$ given by Eq. \eqref{nu-plane} for the
infinite plane. 
Figure \ref{fig1} corresponds to  $2\pi\sigma_0/c<1$, and the red line crosses the blue dashed line at
$x>0$. As $r_0$ increases, the intersection points approach each other and the solutions for the plane and the cylinder
merge.
Figure \ref{fig2} corresponds to $2\pi\sigma_0/c>1$, and the red line crosses the blue solid line also at $x>0$, so 
the relaxation rate in the cylinder geometry is always positive. However it obviously intersects the blue dashed 
line at $x<0$, which corresponds to negative relaxation given by Eq. \eqref{nu-plane}. As $r_0$ increases, the 
actual relaxation rate tends to zero while remaining positive. Hence the straightforward solution of the Drude
equation together with the Maxwell equations for a infinite plane instead of a cylinder of infinite radius gives 
incorrect results.

In \cite{Volkov16}, the negative relaxation rate of plasma oscillations was obtained for $k \ne 0$ and
$2\pi\sigma_0/c \lesssim 1$. However these results were obtained for the case of  a rather strong magnetic field, which smears
the  transition between the  $2\pi\sigma_0/c > 1$ and $2\pi\sigma_0/c < 1$ regimes. In addition to this,
the magnetic field produces a component of current that is uniform in the direction normal to $\bf k$,  in which 
the conductor is also infinite, so  the above reasoning applies to this problem as well. 

The authors of \cite{Volkov16} admit that the electric and 
magnetic fields do not decay with distance from the conducting plane at the point where the relaxation rate
changes sign. This is unphysical and implies that some cutoff parameter should be used. A natural cutoff for 
such a system is its in-plane size, and the correct treatment of this problem should involve a transition from 
a finite-size conductor with appropriate in-plane boundary conditions or topology to an infinite one. This 
would eliminate any unphysical solutions and paradoxical effects.

\end{document}